\documentclass[aps,prb,twocolumn,groupedaddress,showpacs,amsmath]{revtex4}

\usepackage{graphicx}
\usepackage{bm}

\renewcommand{\vec}[1]{\bm{#1}}

\begin{document}

\title{Symmetry breaking in low-dimensional $SU(N)$ antiferromagnets}

\author{Alexei Kolezhuk}
\thanks{On leave from the Institute of Magnetism, Academy of Sciences and
  Ministry of Education, 03142 Kiev, Ukraine}
\affiliation{Institut f\"ur Theoretische Physik C, RWTH Aachen University, 52056 Aachen,
  Germany}
\affiliation{JARA J\"ulich-Aachen Research Alliance, 
Research Centre J\"ulich GmbH, 52425 J\"ulich, Germany} 

\begin{abstract}
Consequences of explicit symmetry breaking in a physically motivated model 
of $SU(N)$ antiferromagnet in spatial dimensions
one and two are studied. 
It is shown that the case $N=3$, which can be realized
in spin-$1$ cold atom systems, displays special properties distinctly different
from those for $N\geq4$. Qualitative form of the phase diagram depending on the
model parameters is given.
\end{abstract}

\maketitle

\section{Introduction} 

During the last several years, there has been a revival of interest
\cite{Nakatsuji+05,Takeya+08,Rizzi+05,Buchta+05,LaeuchliSchmidTrebst06,LegezaSolyom06,TsunetsuguArikawa06,%
LaeuchliMilaPenc06,HaradaKawashimaTroyer07,GroverSenthil07,BarnettTurnerDemler06,Turner+07}
in unconventional types of spin ordering in systems with higher spins
$S\geq 1$. This interest is to large extent motivated by experiments
on Bose-Einstein condensates of cold atoms with internal spin
states.\cite{Stamper-Kurn+98,Schmaljohann+04} Particularly, ordering
of the quadrupole degrees of freedom corresponds to the so-called
``spin-nematic'' type of spin
order,\cite{Papanicolaou88,Chubukov90-91,PodolskyDemler05} which is
difficult to obtain in conventional magnetic materials since its
existence requires the presence of strong non-Heisenberg (biquadratic or
multispin) exchange terms, or the presence of strong frustration
mixing ferro- and antiferromagnetic
couplings.\cite{Shannon+06,Hikihara+08,Sudan+08} 
In strongly frustrated systems with ferromagnetic couplings, 
higher multipolar orders may win over the nematic one, becoming dominant 
correlations.\cite{Momoi+06,Hikihara+08,Sudan+08}
For cold spinful bosons in optical
lattices, strong non-Heisenberg exchange appears in the effective spin
model,
\cite{ZhouSnoek03,ImambekovLukinDemler03} favoring spin-nematic order.
In higher-spin systems, higher symmetries may naturally arise. 
$SU(N)$ generalizations of Heisenberg spin systems in one and two
spatial dimensions have been extensively
studied.\cite{Affleck86-88,ReadSachdev89,ReadSachdev90,ItoiKato97,Assaad05,KawashimaTanabe07}
Several recent studies\cite{GreiterRachel07,CherngRefaelDemler07,Capponi+08}
explore exotic pairing possibilities opened by the existence of higher $SU(N)$
symmetries with $N>2$ in fermionic systems.

In the present paper, we will study what happens to an $SU(N)$
antiferromagnet (AF)
if the high symmetry gets explicitly broken by a weak perturbation.  It will be shown
that, similarly to $N=2$, the physically important case $N=3$ is in many
respects special, and breaking the $SU(3)$ symmetry leads to rich behavior which
might be realizable in cold atom setups. We will see that perturbing the $SU(3)$
symmetry has a drastic effect on the topology, which is
reflected in physical properties due to the role of the Berry phases.
Our starting point will be the $S=1$ model on an anisotropic square
 lattice described by the Hamiltonian
\begin{eqnarray} 
\label{ham1} 
\mathcal{H}&=&\sum_{\vec{n}} \Big\{ \widehat{h}_{\vec{n},\vec{n}+\vec{x}}
+\lambda\widehat{h}_{\vec{n},\vec{n}+\vec{y}} \Big\},\nonumber\\
\widehat{h}_{\vec{n},\vec{n'}}&=&
\cos\theta(\vec{S}_{\vec{n}}\cdot\vec{S}_{\vec{n'}}) 
+\sin\theta(\vec{S}_{\vec{n}}\cdot\vec{S}_{\vec{n'}})^{2}, 
\end{eqnarray}
where $\vec{S}_{\vec{n}}$ is a spin-$1$ operator at the lattice site
$\vec{n}$, and $\lambda$ is the parameter controlling anisotropy of
the lattice, $0<\lambda<1$.  This model appears, particularly, in the
physics of ultracold alkali atoms with hyperfine $S=1$ spins (e.g.,
$^{23}$Na) in optical lattices at odd filling.
\cite{ImambekovLukinDemler03} The parameter $\theta$ can be
varied by tuning the ratio $a_{2}/a_{0}$ of scattering lengths in
$S=2$ and $S=0$ channels using the Feshbach resonance, as well as by creating a
gradient in the optical lattice potential.\cite{Garcia-Ripoll+04}  Similar models
have been proposed \cite{TsunetsuguArikawa06,LaeuchliMilaPenc06} as a
possible explanation for the unconventional spin state discovered
recently \cite{Nakatsuji+05} in the quasi-2d $S=1$ magnet $\rm
NiGa_{2}S_{4}$, and have been also discussed
\cite{HaradaKawashimaTroyer07,GroverSenthil07} in context of the
deconfined quantum criticality conjecture \cite{Senthil+04}.  In one
dimension ($d=1$), this model has been extensively studied and a
number of
analytical\cite{Uimin70Lai74Sutherland75,Johannesson86,Takhtajan82Babujian82-83Kulish+81,%
Parkinson88,Klumper89-90,BarberBatchelor89,AKLT} and
numerical\cite{FathSolyom95,Rizzi+05,Buchta+05,LaeuchliSchmidTrebst06,LegezaSolyom06} results are
available. In two dimensions, it was recently studied numerically by
means of Quantum Monte Carlo technique\cite{HaradaKawashimaTroyer07}
and analytically with the help of a field-theoretical
approach.\cite{GroverSenthil07,NogueiraKragsetSudbo07}

Using the standard representation of the $S=1$ operator
$S_{\vec{n}}^{\alpha}=
-i\epsilon_{\alpha\beta\gamma}t^{\dag}_{\vec{n},\beta}
t^{\vphantom{\dag}}_{\vec{n},\gamma}$
through three bosonic operators $t_{\alpha}$, $\alpha=1,\ldots3$ satisfying the
hardcore constraint
\begin{equation} 
\label{constraint} 
 t^{\dag}_{\alpha}t^{\vphantom{\dag}}_{\alpha}=n_{c}=1,
\end{equation}
one can cast the local Hamiltonian in the form
\begin{eqnarray} 
\label{ham2} 
&&\widehat{h}_{i,j}=
-J t^{\dag}_{i,\alpha}t^{\dag}_{j,\beta}t^{\vphantom{\dag}}_{j,\alpha}
t^{\vphantom{\dag}}_{i,\beta}
-\widetilde{J} t^{\dag}_{i,\alpha}t^{\dag}_{j,\alpha}
t^{\vphantom{\dag}}_{j,\beta}t^{\vphantom{\dag}}_{i,\beta},
\nonumber\\
&& J\equiv -\cos\theta,\quad \widetilde{J}=\cos\theta-\sin\theta.
\end{eqnarray}
Since the model is formulated in terms of local bilinears of bosonic operators
$\vec{t}_{\vec{n}}$, it obviously has the local $U(1)$ symmetry for any values
of the model parameters.  We will be interested in the interval $-3\pi/4 <
\theta < 0$. It is convenient to generalize the Hamiltonian
(\ref{constraint}-\ref{ham2}) by letting the boson flavor index run from $1$ to
$N$ and allowing the parameter $n_{c}$ in (\ref{constraint}) to be an arbitrary
integer number. In case of the related models for cold atoms in optical
lattices, $n_{c}$ has the meaning of the number of atoms per lattice
site,\cite{ZhouSnoek03}  and in what follows we will assume $n_{c}$ to be odd. 
For $n_{c}=1$, $N=3$ corresponds to (\ref{ham1}), $N=2$
describes the spin-$\frac{1}{2}$ XXZ model with $J_{x}=-2(J+\widetilde{J})$,
$J_{z}=2(\widetilde{J}-J)$, and $N=4$ can be realized\cite{LiShen04} as a ``bilayer''
spin-$\frac{1}{2}$ model with four-spin interaction between the
layers:
\begin{eqnarray} 
\label{N=4} 
\widehat{h}_{\vec{n},\vec{n'}}&=&(2\cos\theta-\sin\theta)\big[
  (\vec{s}_{\vec{n}}\cdot\vec{s}_{\vec{n'}}) +
  (\vec{\tau}_{\vec{n}}\cdot\vec{\tau}_{\vec{n'}})\big] \nonumber\\
  &+&4\sin\theta(\vec{s}_{\vec{n}}\cdot\vec{s}_{\vec{n'}})
(\vec{\tau}_{\vec{n}}\cdot\vec{\tau}_{\vec{n'}}), 
\end{eqnarray}
which is essentially the Kugel-Khomskii spin-orbital model
\cite{KugelKhomskii73} with spin and orbital degrees of freedom described by
$\vec{s}_{\vec{n}}$ and $\vec{\tau}_{\vec{n}}$ spin-$\frac{1}{2}$ operators,
respectively; a large number of results are available for this model in one
dimension.\cite{NersesyanTsvelik97,Li+98,Ueda+98,KM98prl,Pati+98,%
Azaria+99,Affleck+99,LeeLee00,Ueda+99,KMS01}

The point $\theta=-\pi/2$ ($J=0$) is remarkable since it has an enhanced
symmetry.  The Hamiltonian (\ref{ham1}) is always $\rm SU(2)$ (generally, $\rm
O(N)$) invariant, but at $\theta=-\pi/2$ the symmetry group is enlarged to $\rm
SU(3)$ (respectively $\rm SU(N)$): since the lattice is bipartite, a
transformation $\vec{t}_{\vec{n}}\mapsto U \vec{t}_{\vec{n}}$ on sites $\vec{n}$
belonging to sublattice $A$ leaves the Hamiltonian invariant if it is
accompanied by a conjugate transformation $\vec{t}_{\vec{n}}\mapsto U^{*}
\vec{t}_{\vec{n}}$ for $\vec{n}\in B$, with a unitary matrix $U$.

Our strategy will be to construct an effective field-theoretical
description of the problem, using $\theta=-\pi/2$ ($J=0$) as a starting
point, and to treat the term proportional to $J$ as a perturbation. We will also
see that a rich behavior is generated by adding another perturbation, 
namely the \emph{easy-axis} single-ion anisotropy
to the $S=1$ Hamiltonian (\ref{ham1}),
\begin{equation} 
\label{hamD} 
\mathcal{H}\mapsto \mathcal{H}-D\sum_{\vec{n}}(S^{z}_{\vec{n}})^{2}, \qquad D>0.
\end{equation}
For the case of a general $N$ this amounts to including the term of
 the form $D\sum_{\vec{n}}t^{\dag}_{\vec{n},N}
 t^{\vphantom{\dag}}_{\vec{n},N}$, which breaks the symmetry down from
 $SU(N)$ to $SU(N-1)$.  In cold atom systems, such terms appear
 naturally in presence of external magnetic field due to the quadratic
 Zeeman effect.\cite{ImambekovLukinDemler03,Zhou+04}

The structure of the paper is as follows: in Sect.\ \ref{sec:efftheory} the
effective continuum theory in the vicinity of the
$SU(N)$-symmetric point is derived, Sect.\ \ref{sec:SUN-ON} considers the
influence of the $\theta$ perturbation breaking the symmetry down to $O(N)$,
Sect.\ \ref{sec:SUN-SUN1} studies the effects of the anisotropy (\ref{hamD}),
and, finally, Sect.\ \ref{sec:summary} contains a brief summary.

\section{Effective field theory \newline in  the vicinity of the  $SU(N)$ point}
\label{sec:efftheory}

To construct the continuum field description,
consider a path-integral representation of the problem,
effectively replacing the bosonic operators $t_{\vec{n},\alpha}$ with
complex fields on the lattice satisfying the constraint in (\ref{constraint}).
To pass to the continuum properly, one should notice
that local spin-quadrupolar correlations are of the ferromagnetic type for
$J+\widetilde{J}>0$, while
the spin-dipolar correlations are 
antiferromagnetic provided $\widetilde{J}>J$;\cite{IvKhymKol08} 
this can be also seen from the numerical results (see Fig.\ 8 of Ref.\ \onlinecite{FathSolyom95}).
We will be
interested mainly in the region of $\theta<0$, where 
the first of
those inequalities is always satisfied, but the second one breaks for
$\theta<\theta_{0}\approx -0.65\pi$.   The theory derived here will be valid for
$\theta>\theta_{0}$, and the proper effective theory for $\theta<\theta_{0}$ can be found in
Ref.\ \onlinecite{IvanovKolezhuk03}.\cite{comment1}

The  AF character of local spin correlations suggests
the following ansatz for the bosonic
lattice fields $\vec{t}_{\vec{n}}$:
\begin{equation} 
\label{ansatz1} 
\vec{t}_{\vec{n}}=(\vec{u}_{n}+i\eta_{\vec{n}}\vec{v}_{\vec{n}})
+(\eta_{\vec{n}}\vec{\varphi}_{\vec{n}} + i\vec{\zeta}_{\vec{n}}),
\end{equation}
where $\eta_{\vec{n}}$ is an oscillating factor taking value $\pm1$ for
$\vec{n}$ belonging to $A$ and $B$ sublattices, respectively,
and $\vec{u}$, $\vec{v}$, $\vec{\varphi}$, and $\vec{\zeta}$ are assumed to be
smooth functions of the site coordinate $\vec{n}$. Defining
  $\vec{z}_{\vec{n}}=(\vec{u}_{\vec{n}}+i\vec{v}_{\vec{n}})/\sqrt{n_{c}}$ and
  $\vec{\psi}_{\vec{n}}=(\vec{\varphi}_{\vec{n}}+i\vec{\zeta}_{\vec{n}})/\sqrt{n_{c}}$, one can
rewrite the above ansatz in a simpler form:
\begin{equation} 
\label{ansatz} 
\vec{t}_{\vec{n}}=\sqrt{n_{c}}\times\begin{cases}
\vec{z}_{\vec{n}}+\vec{\psi}_{\vec{n}}, & \vec{n}\in A \\
\vec{z}_{\vec{n}}^{*}-\vec{\psi}_{\vec{n}}^{*}, & \vec{n}\in B
 \end{cases},
\end{equation}
where the constraints 
\begin{equation} 
\label{constr} 
|\vec{z}|^{2}+|\vec{\psi}|^{2}=1,\quad    
\vec{\psi}\cdot \vec{z}^{*}+\vec{z}\cdot\vec{\psi}^{*}=0
\end{equation}
are implied.
One can expect that the magnitude of $\vec{\psi}$, which corresponds
to ferromagnetic fluctuations, will be much smaller than that of
$\vec{z}$. Using the ansatz (\ref{ansatz}),  passing to the
continuum,  retaining only up to quadratic terms in $\vec{\psi}$ and
neglecting its derivatives, one readily obtains the Euclidean action
$\mathcal{A}= \mathcal{A}_{0}+ \mathcal{A}_{\rm int} +
\mathcal{A}_{B}$, where $\mathcal{A}_{0}$ corresponds to $J=0$:
\begin{eqnarray} 
\label{A0} 
\mathcal{A}_{0}&=&\sqrt{\lambda}n_{c}^{2}\int d\tau \int d^{2}x \Big\{
\frac{1}{n_{c}}(\vec{\psi}^{*}\cdot
\partial_{\tau}\vec{z}-\vec{\psi}\cdot\partial_{\tau}\vec{z}^{*})
 \nonumber\\
&+&4\widetilde{J}(1+\lambda)\big[|\vec{\psi}|^{2}-|\vec{\psi}^{*}\cdot\vec{z}|^{2}\big]\nonumber\\
&+&\widetilde{J}\big[
  |\partial_{k}\vec{z}|^{2}
-|\vec{z}^{*}\cdot\partial_{k}\vec{z}|^{2} 
 \big] \\
&+&\mu_{1}(\vec{\psi}\cdot\vec{z}^{*}+\vec{\psi}^{*}\cdot\vec{z})
+\mu_{2}(|\vec{z}|^{2}+|\vec{\psi}|^{2}-1)\Big\},\nonumber
\end{eqnarray}
and $\mathcal{A}_{B}$ is the topological Berry phase contribution 
\begin{equation} 
\label{AB} 
\mathcal{A}_{B}=in_{c}\sum_{\vec{n},\tau} \eta_{\vec{n}} \arg\big(\vec{z}^{*}_{\vec{n}}(\tau)\cdot\vec{z}_{\vec{n}}(\tau+d\tau)\big)
\end{equation}
which is known to play a crucial
role in the physics of the system.\cite{Haldane88,ReadSachdev89,ReadSachdev90} 
It is important to realize\cite{GroverSenthil07,ZhouSnoek03} that the naive
continuum limit of (\ref{AB}),
\begin{equation}
\label{ABcont}
\mathcal{A}_{B}=n_{c}\int d\tau \sum_{\vec{n}}
\eta_{\vec{n}}\vec{z}^{*}_{\vec{n}}\cdot\partial_{\tau}\vec{z}_{\vec{n}},
\end{equation}
can only capture the contributions from smooth field
configurations and in case of dominant nematic correlations, when $\vec{z}$
becomes a real vector defined up to a sign, misses the additional phase stemming from
disclinations.

The term  $\mathcal{A}_{\rm int}$ is determined by the ``perturbation'' $J$,
\begin{eqnarray} 
\label{Aint} 
\mathcal{A}_{int}&=&J\sqrt{\lambda}n_{c}^{2}\int d\tau \int d^{2}x \Big\{ 
-|\vec{z}^{2}|^{2} +|\vec{z}\cdot
  \partial_{k}\vec{z}|^{2} \nonumber\\
&+&\big[
  \vec{z}^{2}{\vec{\psi}^{*}}^{2}
  +\frac{1}{2}\vec{z}^{2}(\partial_{k}\vec{z}^{*})^{2} +\mbox{c.c.}\big]
\Big\}.
\end{eqnarray}
Here the index $k$ runs over
two spatial coordinates, the factor $\sqrt{\lambda}$ in
(\ref{A0}), (\ref{Aint}) comes from rescaling one of those
coordinates to compensate for the anisotropy of interactions, and
$\mu_{1,2}$ are the Lagrange multipliers ensuring the constraints.

For $J=0$, one can easily integrate out $\vec{\psi}$ and $\mu_{1}$
fields; it turns out that 
$\mu_{1}=-n_{c}^{-1}\vec{z}^{*}\cdot\partial_{\tau}\vec{z}$, which yields
\begin{eqnarray} 
\label{psi} 
\vec{\psi}&=&-[4\widetilde{J}n_{c}
(1+\lambda)]^{-1}\big\{\partial_{\tau}\vec{z}-\vec{z}(\vec{z}^{*}\cdot
\partial_{\tau}\vec{z}) \big\},\nonumber\\
\vec{\psi}^{*}&=&[4\widetilde{J}n_{c}
(1+\lambda)]^{-1}\big\{\partial_{\tau}\vec{z}^{*}+\vec{z}^{*}(\vec{z}^{*}\cdot
\partial_{\tau}\vec{z}) \big\}.
\end{eqnarray}
Substituting this back into (\ref{A0}), one obtains the
effective action for $\vec{z}$ field only, where we can now
approximately assume $|\vec{z}|^{2}=1$. 
Rescaling the imaginary time axis $\tau\mapsto
\tau/(2n_{c}\widetilde{J}\sqrt{1+\lambda})$, one arrives at the
effective action 
\begin{equation} 
\label{CP-N}
 \mathcal{A}_{0}=\frac{1}{2g}\int d^{d+1}x \Big\{
 |\partial_{\mu}\vec{z}|^{2} -|\vec{z}^{*}\cdot\partial_{\mu}\vec{z}|^{2}
\Big\},\quad
g=\frac{\sqrt{1+\lambda^{-1}}}{n_{c}},
\end{equation}
where $d=2$ is the spatial dimension, and the index $\mu$ runs over
all $d+1$ space-time coordinates. Had we started with a single $S=1$ chain
instead of the square lattice, we would have obtained the action of the same
form (\ref{CP-N}), but with $d=1$ and $g=1/n_{c}$.
This  is nothing but the action of the 
$\rm CP^{N-1}$
model,\cite{Eichenherr78,GoloPerelomov78,DAdda+78,Witten79,ArefevaAzakov80}
originally proposed as an effective
theory for $\rm SU(N)$ antiferromagnets by  Read and
Sachdev \cite{ReadSachdev89,ReadSachdev90}. This action has a local $\rm
U(1)$ gauge symmetry $\vec{z}\mapsto e^{i\varphi(x)} \vec{z}$ and can be
rewritten in the form
\begin{equation} 
\label{CP-N1} 
\mathcal{A}_{0}=\frac{1}{2g}\int d^{d+1}x |(\partial_{\mu}-iA_{\mu})\vec{z}|^{2},
\end{equation}
where $A_{\mu}=i(\vec{z}\cdot\partial_{\mu}\vec{z}^{*})$ is the $\rm U(1)$
gauge field.

The $CP^{N-1}$ model without the topological phase term is always gapped in
$d=1$ and displays an ordering transition in $d=2$ at a certain critical value
of the coupling constant.\cite{DAdda+78,Witten79}
In the disordered phase the $\vec{z}$ field acquires a finite mass, and 
a kinetic term for
the gauge field is dynamically generated, \cite{Witten79}
\begin{equation}
\label{gauge-kin}
\mathcal{A}\mapsto \mathcal{A} +
 \frac{N}{4e_{0}^{2}} \int d^{d+1}x  F_{\mu\nu}^{2},
\end{equation} 
where $F_{\mu\nu}=\partial_{\mu}A_{\nu}-\partial_{\nu}A_{\mu}$, and the coupling
constant $e_{0}^{2}\propto \Delta^{3-d}$.

The Berry phase term (\ref{AB}) is crucial for the physics of the disordered
phase; \cite{Haldane88,ReadSachdev89,ReadSachdev90} particularly, it leads to
spontaneous dimerization in $d=1$ for odd $n_{c}$ (except for $N=2$ which is
special: in that case the system remains gapless and translationally invariant
in a wide $g$ range\cite{Haldane83,ShankarRead90,Affleck91-prl,Azcoiti+07}), and
in two dimensions the disordered phase gets spontaneously dimerized in different
patterns depending on the value of $(n_{c} \mod 4)$.  We will come back to the
role of the Berry term later and look into the rest of the action first.

\section{Effect of the $\rm SU(N)\mapsto O(N)$ perturbation}
\label{sec:SUN-ON}

The perturbing action (\ref{Aint}) explicitly breaks the global
$\rm SU(N)$ symmetry down to $\rm O(N)$, but preserves the $\rm U(1)$ gauge
symmetry. Consequently, nonzero $J$ can produce only gauge-invariant
perturbation terms of the form
\[
|\vec{z}^{2}|^{2},\quad |\vec{z}\cdot
D_{\mu}\vec{z}|^{2} ,\quad\ldots,
\]
where $D_{\mu}\equiv \partial_{\mu}-iA_{\mu}$ and the ellipsis stands for terms
with higher derivatives. It is easy to see that the first term above is relevant
for $d<3$, while the second one is irrelevant for $d>1$ (for $d=1$ it is
marginal). It thus makes sense to consider only the effect of the most relevant
term, which brings us to the perturbed action $\mathcal{A}_{\gamma}=
\mathcal{A}_{0}+\mathcal{A}_{\rm int}$, with
\begin{equation} 
\label{Apert}
\mathcal{A}_{\rm int}= -\frac{\gamma}{2g}\int d^{d+1}x
|\vec{z}^{2}|^{2}, \qquad \gamma\simeq \frac{J}{\widetilde{J}}.
\end{equation}

It is easy to generalize the standard large-$N$ mean-field analysis \cite{DAdda+78} of
the $\rm CP^{N-1}$ model to include the effect of the $\rm SU(N)$-breaking
perturbation $\gamma$. We consider the action 
\begin{eqnarray} 
\label{A-saddle}
 \mathcal{A}_{MF}&=&\frac{1}{2g}\int d^{d+1}x \Big\{ 
 |\partial_{\mu}\vec{z}|^{2} -|\vec{z}^{*}\cdot\partial_{\mu}\vec{z}|^{2}\nonumber\\
&-&\gamma |\vec{z}^{2}|^{2}+\sigma (|\vec{z}|^{2}-1)\Big\},
\end{eqnarray}
where $\sigma$ is the Lagrange multiplier responsible for the constraint
$|\vec{z}|^{2}=1$, 
and expand it around a stationary saddle-point solution $\vec{z}=\vec{z}_{0}$,
$\sigma=\sigma_{0}$. This expansion has to be performed differently depending on
whether the perturbation is of the ``nematic'' ($\gamma>0$) or ``antiferromagnetic''
($\gamma<0$) type.

\subsection{``Nematic'' side ($\gamma >0$)}

In this case the saddle point  can be chosen in the form
$\vec{z}_{0}=(n_{0},0,\ldots,0)$, which in our original $N=3$ model corresponds
to the spin-nematic (quadrupolar) order.  Fluctuations around the mean-field
solution, $\vec{z}=\vec{z}_{0}+\vec{u}+i\vec{v}$ can be described by two real
$N$-component vectors $\vec{u}$, $\vec{v}$. Due to the constraint
$|\vec{z}|^{2}=1$ one can set $u_{1}\approx 0$, and the gauge-fixing condition
(e.g., setting $n_{0}$ to be real) yields $v_{1}\approx 0$. After integration
over quadratic fluctuations, the saddle point equations are obtained as
\begin{eqnarray} 
\label{mf-nem}
&& n_{0}^{2}+g(N-1)\sum_{\vec{k}} \Big\{ 
\frac{1}{\sigma_{0}+k^{2}} +\frac{1}{\sigma_{0}+4\gamma +k^{2}}
\Big\}=1,\nonumber\\
&& \sigma_{0}n_{0}=0,
\end{eqnarray}
where the sum is over $(d+1)$-dimensional reciprocal space.
In one spatial dimension, $d=1$, the model is disordered ($n_{0}=0$) for any
value of the coupling constant $g$, and the field $\vec{z}$ is always massive,
with $\sigma_{0}=\Delta^{2}$ having the meaning of the squared spectral gap,
\begin{equation} 
\label{sigma-1d-nem}
 \Delta^{2}\simeq \Lambda^{2}\exp\Big\{-\frac{2\pi}{g(N-1)} \Big\} -2\gamma,
 \quad \gamma \ll
\Delta^{2},
\end{equation}
where $\Lambda$ is the lattice (UV) cutoff and it is assumed that $\gamma \ll
\Delta^{2}$ and both $\gamma$ and $\Delta$ are small compared to the cutoff. In
the opposite case if $\gamma \gg
\Delta^{2}$ one obtains
\begin{equation} 
\label{sigma-1d-nem-b}
 \Delta\simeq \frac{\Lambda^{2}}{2\sqrt{\gamma}}\exp\Big\{-\frac{2\pi}{g(N-1)} \Big\} ,
 \quad \gamma \gg
\Delta^{2}.
\end{equation}

In two dimensions, $d=2$,
there is a finite second-order transition point
$g=g_{c}$, given by 
\begin{equation} 
\label{gc-nem} 
g_{c}^{-1}\simeq\frac{N-1}{\pi^{2}}(\Lambda -\frac{\pi}{2}\sqrt{\gamma}),
\end{equation}
such that for $g<g_{c}$ the $O(N)$ symmetry is spontaneously broken and the
ground state is ordered, $n_{0}^{2}=1-g/g_{c}$, while for $g>g_{c}$ one has a
disordered phase with $n_{0}=0$ and $\sigma_{0}=\Delta^{2}$, where the gap
$\Delta$ behaves as $\Delta\simeq \frac{4\pi}{N-1}(g_{c}^{-1}-g^{-1})$ 
at $g\to g_{c}$.

The transition at $g=g_{c}$ corresponds in our original model (\ref{ham1}) to a
transition at some critical value of anisotropic coupling $\lambda=\lambda_{c}$,
so that one has the spin-nematic ordered phase at $\lambda>\lambda_{c}$ and the
quantum disordered phase at $\lambda<\lambda_{c}$, and 
the ``disordered'' phase actually corresponds to a dimerized state arising due
to the Berry phase term.\cite{ReadSachdev90}  The critical value $\lambda_{c}$ can be estimated
using known large-$N$ result\cite{ReadSachdev90} for the critical point at
$\gamma=0$ and the isotropic square lattice ($\lambda=1$), $n_{c}^{\rm
crit}\approx 0.19 N$, which yields $\lambda_{c}^{-1}(\gamma=0)\approx 55.4
(n_{c}/N)^{2}-1$. Since only $0\leq\lambda_{c}\leq 1$ makes sense, the latter
estimate suggests that in absence of the perturbation $\gamma$ the system does
not order for any value of $\lambda$ for $N>N_{c}\simeq 5$.  Extrapolating to
$N=3$, one obtains that at $\gamma=0$ (i.e., $\theta=-\pi/2$) the
critical coupling is $\lambda_{c}\approx 0.19$, while the Quantum Monte Carlo
calculations\cite{HaradaKawashimaTroyer07} done for bilinear-biquadratic $S=1$
model yield $\lambda_{c}\simeq 0.13$.

\begin{figure}[tb]

\includegraphics[width=0.4\textwidth]{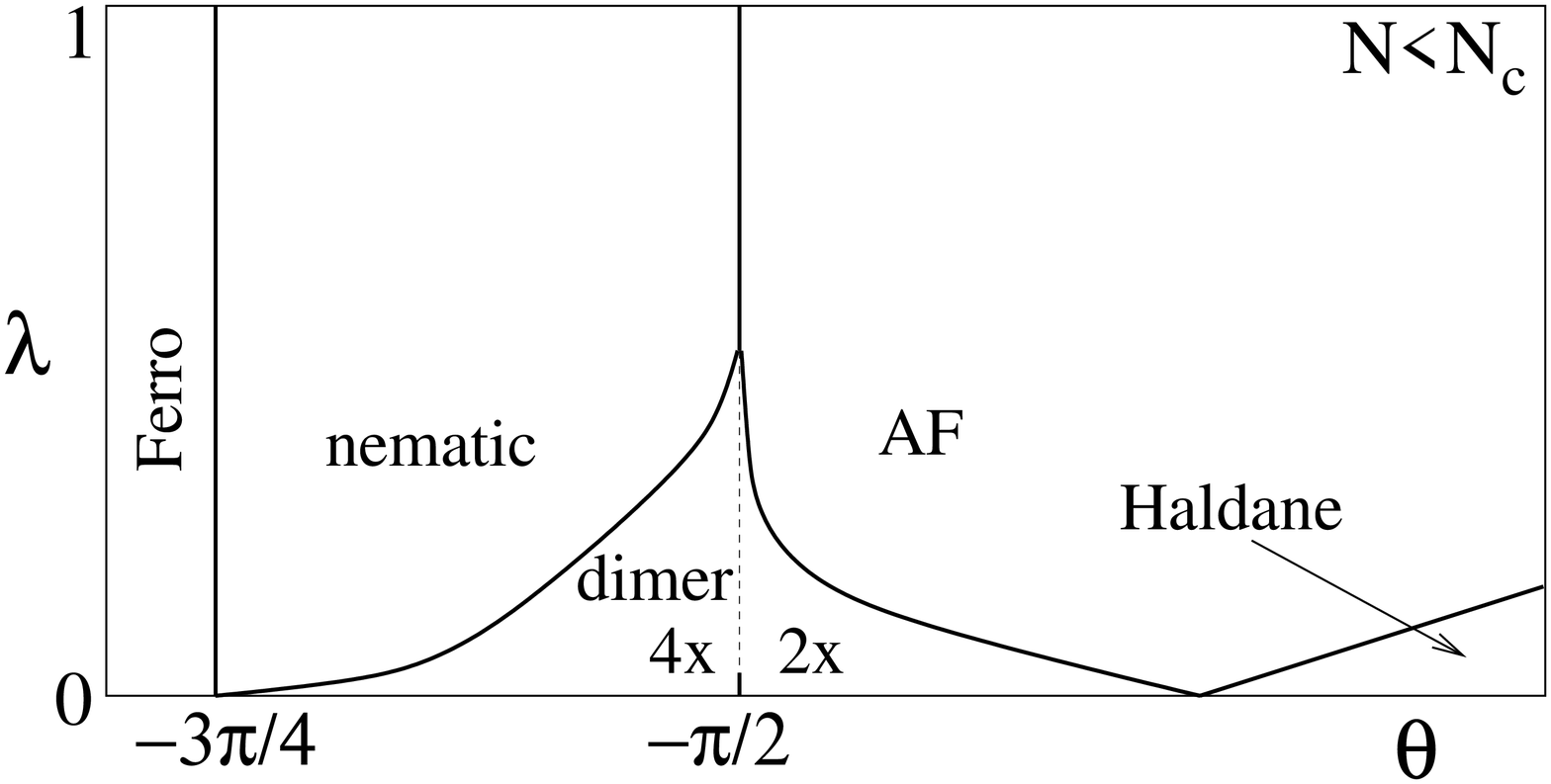}
\caption{\label{fig:LaTh-diag}
A sketch of the phase diagram of the model (\ref{ham2}) on an anisotropic square
lattice in the vicinity of the $SU(N)$-symmetric point $\theta=-\pi/2$. For
$N<N_{c}\simeq 5$ the
phase boundary has a square-root-type cusp at $\theta\to -\pi/2$ as suggested by
Eqs.\ (\ref{gc-nem}), (\ref{gc-af}). For $N>N_{c}$ the dimerized phase has a
finite extent at $\lambda=1$. The phase denoted as ``Haldane'' is for $N=3$ indeed the Haldane
phase whose boundary lies at $\theta=-\pi/4$, and in case of $N=4$ it is the staggered dimer
phase\cite{NersesyanTsvelik97,KM98prl} with the boundary at $\theta=0$. For
$N=3$ only, the
degeneracy of the
dimerized phase is twofold for $\theta>-\pi/2$ and fourfold for $\theta<-\pi/2$,
see Sect.\ \ref{subsec:Berry}.
}
\end{figure}

\subsection{``Antiferromagnetic'' side ($\gamma<0$)}

For $\gamma<0$ the interaction favors the minimal absolute value of
 $\vec{z}^{2}$, so the saddle-point solution can be chosen in the form
 $\vec{z}_{0}=2^{-1/2}(n_{0},in_{0},\ldots,0)$, which in the original $N=3$
 model corresponds to a finite AF order parameter
 $\vec{\ell}=-i(\vec{z}_{0}^{*}\times \vec{z}_{0})$. For the fluctuations
 $\vec{u}$, $\vec{v}$ one can set $u_{2}\approx 0$, $v_{1}\approx 0$ to fix the
 gauge, and the constraint $|\vec{z}|^{2}=1$ yields $u_{1}+v_{2}\approx 0$. The
 mean field equations take the form
\begin{eqnarray} 
\label{mf-af}
&& n_{0}^{2}+2g\sum_{\vec{k}} \Big\{ 
\frac{N-2}{\sigma_{0}+k^{2}} +\frac{1}{\sigma_{0}+4|\gamma| +k^{2}}
\Big\}=1,\nonumber\\
&& \sigma_{0}n_{0}=0. 
\end{eqnarray}
For $d=1$ there is again only a
disordered phase with $n_{0}=0$ and $\sigma_{0}=\Delta^{2}$, where
\begin{equation} 
\label{sigma-1d-af} 
\Delta^{2}\simeq\Lambda^{2}\exp\Big\{-\frac{2\pi}{g(N-1)} 
\Big\} -\frac{4|\gamma|}{N-1}
\end{equation}
under the assumption $|\gamma|\ll \Delta^{2}$, and in case $\Delta^{2}\gg
|\gamma|$ the gap is given by
\begin{equation} 
\label{sigma-1d-af-b} 
\Delta\simeq\Lambda\Big(
\frac{\Lambda^{2}}{4|\gamma|}\Big)^{\frac{1}{2(N-2)}}
\exp\Big\{-\frac{\pi}{g(N-2)} 
\Big\}.
\end{equation}

For $d=2$ the system orders at $g$ below the critical value $g_{c}$ given by
\begin{equation} 
\label{gc-af} 
g_{c}^{-1}\simeq\frac{\Lambda(N-1)}{\pi^{2}}  -\frac{\sqrt{|\gamma|}}{\pi},
\end{equation}
and for $g>g_{c}$ one has a disordered phase with a finite gap
$\Delta=\sqrt{\sigma_{0}}$ which grows linearly in the vicinity  of the
transition, $\Delta\simeq \frac{2\pi}{N-2}(g_{c}^{-1}-g^{-1})$. 

The corresponding phase diagrams are sketched in Fig.\ \ref{fig:LaTh-diag}.  On
the AF side the effect of perturbation $\gamma$ is weaker by a factor of $\sim
1/N$ compared to the ``nematic'' case $\gamma>0$: for $d=2$ this is translated
into different amplitudes of the square-root cusp in the dependence of the
critical coupling $\lambda_{c}$ on $\gamma$  for $\gamma>0$ ($\theta<-\pi/2$) and
$\gamma<0$ ($\theta>-\pi/2$).  For $d=1$ this effect should be seen in different
slopes of the gap $\Delta(\gamma)$ for positive and negative
$\gamma$; this is in line with the results from exact diagonalization
of small finite chains\cite{FathSolyom93} as well
as with the recent density matrix renormalization group calculations for the
model (\ref{ham1}) on a
ladder.\cite{LaeuchliSchmidTrebst06}

\subsection{Influence of the perturbation on the Berry term}
\label{subsec:Berry}

Up to now we have considered only the effect of the $SU(N)$-breaking
perturbation $\gamma$ on the action without the Berry term. Apart from
favoring  nematic or antiferromagnetic order, the effect consists in a mere
shift  of the
transition point in two dimensions, and a change of the gap in $d=1$ case.
However, there is another important effect of the perturbation $\gamma$: 
as we will see, it drastically affects the Berry term, which has important
consequences for the physics of the disordered phase.

\subsubsection{$d=1$}

The role of the Berry phase term $\mathcal{A}_{B}$ at $\gamma=0$ has
been studied in detail.\cite{Affleck86-88,ReadSachdev90,Affleck91-prl}
In the one-dimensional case one obtains
\begin{equation}
\label{AB-1d}
 \mathcal{A}_{B}^{(d=1)}=i\Theta q, \qquad 
 q=\frac{1}{2\pi}\int dx\,d\tau\,F_{x\tau}, 
\end{equation}
where the integer number $q$ 
has the meaning of the net topological charge (skyrmion number), and 
\[
\Theta=(\pi n_{c} \bmod 2\pi)
\]
is the so-called
topological angle. Explicitly expressed through $\vec{z}$, the topological
charge reads
\begin{equation} 
\label{q-CPN} 
q=-\frac{i}{2\pi}\int d^{2}x \epsilon_{\mu\nu}(\partial_{\mu}\vec{z}^{*}\cdot
\partial_{\nu}\vec{z}).
\end{equation}
For even $n_{c}$ the Berry
phase has no effect, while for odd $n_{c}$ 
it leads to the twofold
degenerate ground state
with a 
finite  
``static electric
field'' (topological charge density) 
\begin{equation} 
\label{Efield1} 
\langle iF_{x\tau}\rangle=\pm\frac{e_{0}^{2}}{N}.
\end{equation}

One can easily show that the topological charge density is directly
proportional to the dimerization order parameter, in essentially the
same way as it has been done \cite{ShankarMurthy05} for the $O(3)$
nonlinear sigma model. Indeed, the dimerization operator at $\gamma=0$
can be defined as
\begin{equation} 
\label{Odim-def} 
\mathcal{O}^{\rm dim}_{n} = \eta_{n}(\vec{S}_{n}\cdot\vec{S}_{n+1})^{2}
\end{equation}
and after passing to the continuum its leading non-oscillating part will take
the form
\begin{eqnarray} 
\label{Odim} 
\mathcal{O}^{\rm dim}_{n} &\mapsto& (\vec{\psi}_{n}\cdot \vec{z}_{n+1}^{*}
-\vec{z}_{n}\cdot \vec{\psi}_{n+1}^{*}) + \mbox{c.c.}\nonumber\\ 
&\mapsto&
2(\vec{\psi}\cdot \partial_{x}\vec{z}^{*} +\vec{\psi}^{*}\cdot
\partial_{x}\vec{z})\nonumber\\ &\mapsto& [2\widetilde{J}
  n_{c}(1+\lambda)]^{-1}(\partial_{\tau}\vec{z}^{*}\cdot \partial_{x}\vec{z}
-\partial_{\tau}\vec{z}\cdot \partial_{x}\vec{z}^{*} )\nonumber\\ 
&=&
[2\widetilde{J} n_{c}(1+\lambda)]^{-1} (iF_{x\tau})
\end{eqnarray}
Thus, at the $SU(N)$-symmetric point $\gamma=0$ the ground state for
odd $n_{c}$ and $N\geq 3$ is spontaneously
dimerized.\cite{ReadSachdev90} The case $N=2$, however, is an
exception: for $N=2$ the model is equivalent to the $O(3)$ nonlinear
sigma model with the topological angle $\Theta=\pi$, which is gapless
and nondimerized \cite{ShankarRead90,Affleck91-prl,Azcoiti+07} in a
wide range of the coupling $g$.

Let us first illustrate the effect of the $SU(N)\mapsto O(N)$ perturbation
$\gamma$ on the Berry phase by a simple observation\cite{IvKhymKol08}
valid for
$N=3$. Finite $\gamma<0$ favors field configurations of the antiferromagnetic
type, namely $\vec{z}=\frac{1}{\sqrt{2}}(\vec{e}_{1}+\vec{e}_{2})$, with
$\vec{e}_{1,2}$ being two orthonormal vectors and
$\vec{n}(\theta,\varphi)=\vec{e}_{1}\times \vec{e}_{2}$ having the meaning of
the unit N\'eel vector characterized by two spherical angles $\theta$ and
$\varphi$. It is a straightforward exercise to check that
\begin{eqnarray} 
\label{q=2Q}
q&=&\frac{1}{2\pi}\int d^{2}x\sin\theta
\epsilon_{\mu\nu}(\partial_{\mu}\theta)( \partial_{\nu}\varphi)\nonumber\\
&=& =\frac{1}{4\pi}\int d^{2}x \epsilon_{\mu\nu}\vec{n}\cdot
(\partial_{\mu}\vec{n}\times \partial_{\nu}\vec{n}) =2Q,
\end{eqnarray}
where the topological charge $Q$ is the winding number of the $S^{2}\mapsto
S^{2}$ mapping characterizing the space-time distribution of the unit vector
$\vec{n}(\theta,\varphi)$. This shows that negative $\gamma$ favors
$\vec{z}$-field configurations with \emph{even} charge $q$ and suggests that
configurations with odd $q$ become suppressed. This is physically important,
because if odd-$q$ configurations are prohibited, the Berry term obviously
becomes ineffective, irrespectively of whether $n_{c}$ is even or odd. The above
argument cannot be applied for $N>3$ because the second homotopy group of
$O(N>3)$ sigma models is trivial so they possess no $\pi_{2}$ topological
charge. It is also not possible to extend this argument to $\gamma>0$, because
in this case ``nematic'' configurations with $\vec{z}$ being a real (up to an
arbitrary overall phase) unit vector are favored and for such configurations the
$CP^{N-1}$ topological charge (\ref{q-CPN}) identically vanishes.

To understand what happens in case of general $N$ and $\gamma$, consider the
general one-skyrmion ($q=1$) solution of the $1+1$-dimensional $CP^{N-1}$ model
which has the form
\begin{equation} 
\label{q1-gen} 
z_{\alpha}=\frac{c_{\alpha}(Z-a_{\alpha})}
{\Big(\sum_{\beta}|c_{\beta}|^{2}|Z-a_{\beta}|^{2}\Big)^{1/2}},
\end{equation}
where $Z=x_{0}+ix_{1}$ is the complex coordinate, the complex numbers
$a_{\alpha}$ have the meaning of coordinates of the $N$ skyrmion constituents
(sometimes called ``zindons'' from a Persian word meaning prison
\cite{DiakonovMaul00}), and another set of complex numbers $c_{\alpha}$ may be
viewed as amplitudes associated with each zindon. Normalizing the amplitudes
$c_{\alpha}$ as $\sum_{\alpha}|c_{\alpha}|^{2}=1$, putting the origin into the
``center of mass'' (which amounts to demanding
$\sum_{\alpha}|c_{\alpha}|^{2}a_{\alpha}=0$), and defining the average ``size''
$R$ of the skyrmion as the dispersion of the zindon positions,
\begin{equation} 
\label{skyrmion-size} 
R^{2}\equiv \sum_{\alpha}|c_{\alpha}|^{2}|a_{\alpha}|^{2},
\end{equation}
one can recast the general $q=1$ solution (\ref{q1-gen}) in a more elegant form
\cite{DAdda+78}
\begin{equation} 
\label{q1-nice} 
\vec{z} =\frac{\vec{U}R+\vec{V} Z}{\big(|Z|^{2}+R^{2}\big)^{1/2} },
\end{equation}
where  $\vec{U}$, $\vec{V}$ are two orthonormal complex $N$-component vectors,
\begin{equation} 
\label{UV} 
\vec{U}^{*}\cdot\vec{U}=\vec{V}^{*}\cdot\vec{V}=1, \quad \vec{U}^{*}\cdot\vec{V}=0.
\end{equation}
For $\gamma=0$, i.e. in the unperturbed $CP^{N-1}$ model, the action of such 
skyrmion solution does not depend on its parameters. For a finite $\gamma$,
however, one gets an additional contribution to the action from  the
$|\vec{z}^{2}|^{2}$ term (\ref{Apert}). 

Let us calculate this correction to the first order in $\gamma$.  Consider first
the ``antiferromagnetic'' case $\gamma<0$.  To minimize the action cost, we must
reduce as much as possible the deviations of $\vec{z}^{2}$ from $0$. Requiring
$\vec{U}^{2}=0$ ensures that $\vec{z}^{2}\to 0$ at $|Z|\to \infty$, killing the
next leading term in $Z$ fixes $\vec{U}\cdot\vec{V}=0$, and, finally, if we were
able to satisfy additionally $\vec{V}^{2}=0$, then the condition $\vec{z}^{2}=0$
would be identically fullfilled. Those three constraints can be satisfied
together with (\ref{UV}) only if the four real $N$-component vectors
$\mbox{Re}(\vec{U})$, $\mbox{Re}(\vec{V})$, $\mbox{Im}(\vec{U})$,
$\mbox{Im}(\vec{V})$ are mutually orthogonal, which is readily achieved for
$N\geq 4$ but is obviously impossible for $N=3$. Thus, \emph{for $N\geq 4$ the
$q=1$ skyrmion (\ref{q1-nice}) remains an exact solution even for finite
$\gamma<0$}.  In other words, when $\gamma<0$ is switched on, the ``zindons''
constituting a skyrmion are able to adjust themselves for $N\geq 4$ in such a
way that the skyrmion continues to provide the minimum of action.  This is in
fact amusing because formally for $\gamma<0$ the model has only the $O(N)$
symmetry, and one would expect that skyrmions do not exist for $N>3$.

For $N=3$, $\gamma<0$ the minimum contribution of the
perturbation to the action of the skyrmion (\ref{q1-nice}) is achieved if
$\vec{V}$ is real and the three vectors $\vec{V}$, $\mbox{Re}(\vec{U})$, and
$\mbox{Im}(\vec{U})$ are mutually orthogonal. The excess action due to finite
$\gamma$ is then given by
\begin{equation} 
\label{dA-af}
\Delta{\mathcal A}_{\gamma<0}=-\frac{\gamma}{2g}\int d^{2}x
|\vec{z}^{2}|^{2} = -\frac{\pi \gamma R^{2}}{2g},
\end{equation}
and it grows as a square of the skyrmion size which means at $\gamma<0$ the
field configurations with $q=1$ are prone to collapse and only exist as
metastable ``excitations''. At the same time, one can easily adjust the
parameters of a general $q=2$ skyrmion solutions of the $\gamma=0$ model so that
$\vec{z}^{2}=0$ is identically satisfied (see Appendix \ref{app:A}). This effect
can be interpreted as ``topological pairing'' of $q=1$
skyrmions.\cite{IvKhymKol08} The total topological charge density $F_{x\tau}$
can be separated into two parts, $F_{x\tau}^{(q=1)}$ and $F_{x\tau}^{(q=2)}$
which correspond to the contributions from unbound $q=1$ skyrmions and their
bound pairs, respectively. Only $F_{xt}^{(q=1)}$ contributes to the nontrivial
part of the Berry phase (\ref{AB-1d}), while the full $F_{x\tau}$ enters the
gauge field kinetic energy term (\ref{gauge-kin}).  The dimerization order
parameter $\langle \mathcal{O}^{\rm dim}\rangle \propto \langle
iF_{x\tau}\rangle$ will be proportional to the fraction $\rho$ of the $q=1$
skyrmions and so will be diminishing with increasing $|\gamma|$. This
``topological'' suppression of dimerization at $\gamma<0$ exists only for $N=3$
and is absent for $N\geq 4$, which implies that within our description the
dimerized phase  for $N\geq4$ should extend to the entire region $\gamma<0$ (i.e., up to
$\theta=\pi/4$ which in our notation is another $SU(N)$-symmetric point of the
model, corresponding to the transition into a critical phase\cite{Uimin70Lai74Sutherland75});
however, it is clear that our description will eventually break down as the ``perturbation''
$|\gamma|$ becomes large. In fact, according to exact results (see Ref.\
\cite{Tu+08}) for one-dimensional $SO(N)$ generalizations of the
bilinear-biquadratic model (\ref{ham1}), even for $N\geq 4$ there is still a
phase transition on the way from the AF $SU(N)$ point to the critical $SU(N)$
one.  The chunk of different phase lying between the dimerized phase and the
critical $SU(N)$ point gets squeezed with increasing $N$, and the transition
point for $N\geq 4$ lies in the region of $\gamma\gtrsim 1$, way beyond the
range of applicability of the present approach.

Consider now the perturbation of the opposite sign $\gamma>0$, which favors
nematic-like field configurations with $\vec{z}=e^{i\alpha}\vec{\varphi}$, where
$\alpha$ is an arbitrary phase and $\vec{\varphi}$ is a real unit vector. For
such $\vec{z}$ the topological charge (\ref{q-CPN}) is identically zero, which
indicates that skyrmions with any charge are suppressed by the perturbation. In
a different way one can see that by calculating the $\gamma$-dependent
correction to the action.  For a $q=1$ skyrmion (\ref{q1-nice}) minimizing the
deviation of $|\vec{z}^{2}|$ from $1$ leads to the requirement that $\vec{U}$
and $\vec{V}$ are real, and the resulting correction diverges logarithmically
with the system size $L$,
\begin{equation} 
\label{dA-nem} 
\Delta{\mathcal A}_{\gamma>0} \simeq \frac{2\pi \gamma R^{2}}{g}\ln\frac{L}{R}.
\end{equation}
In the disordered phase, one expects that the system size $L$ above
will be replaced by the correlation length $\xi$.  In contrast to the
AF-like case $\gamma<0$, this suppression persists for any number of
the field components $N$.  A similar calculation for $q=2$ yields
$\Delta \mathcal{A}^{q=2}_{\gamma >0} \propto \gamma R^{2}$, so the
even-charged skyrmions are suppressed as well, though weaker than the
odd-charged ones.  Thus, with increasing $\gamma$ the contribution
from smooth field configurations (skyrmions) to the Berry phase dies out, but at the same
time the contribution from discontinuous configurations
(disclinations, or $Z_{2}$ vortices) remains unaffected and gradually
becomes the leading one.  Indeed, a configuration with a real vector
$\vec{z}$ abruptly changing sign across some bond along a path running
in the time direction contributes the Berry phase equal to $\pi$ for
every such bond,\cite{ZhouSnoek03,GroverSenthil07} which is not
captured by the continuum-limit expression (\ref{AB-1d}) but is
readily seen from the general formula (\ref{AB}). In the disordered
phase the fluctuations of $\vec{z}$ are gapped and can be integrated
out, leaving one only with Ising-like degrees of freedom marking bonds
where a discontinuous change $\vec{z}\to -\vec{z}$
occurred.\cite{GroverSenthil07} The resulting so-called odd $Z_{2}$
gauge theory \cite{SenthilFisher00} is known to be always dimerized in
one dimension,\cite{MoessnerSondhiFradkin02} which, according to Grover and
Senthil,\cite{GroverSenthil07} explains why the dimerized phase extends all the
way up to $\theta=-3\pi/4$ (their arguments can be literally transferred to the
effective theory of Ref.\ \onlinecite{IvanovKolezhuk03} which is suited for
describing the region $-3\pi/4 < \theta \lesssim -0.65\pi$ with ferro-type local correlations).

\subsubsection{$d=2$}

In two dimensions the Berry phase is determined by instanton processes
(``monopoles'') changing the skyrmion topological quantum number $q$ and is
given by \cite{Haldane88,ReadSachdev90}
\begin{equation} 
\label{AB-2d} 
\mathcal{A}_{B}=\frac{i\pi n_{c}}{2}\sum_{\vec{r}_{i}} \zeta(\vec{r}_{i}) 
\widetilde{q}_{i},
\end{equation}
where the sum is over the locations $\vec{r}_{i}$ of monopoles having the charge
$ \widetilde{q}_{i}$ (i.e., the skyrmion number gets changed by
$\widetilde{q}_{i}$), and the factor $\zeta(\vec{r}_{i})$ takes values $0$, $1$,
$2$, $3$ for $\vec{r}_{i}$ belonging to the four dual sublattices $W$, $X$, $Y$,
$Z$ respectively (see Fig.\ 7 of Ref.\ \onlinecite{ReadSachdev90}).  At the
$SU(N)$-symmetric point $\gamma=0$ for $n_{c}\not=0 \mod 4$
the Berry term leads to the ground state with
nonzero instanton density, thus to finite
electric fields and to spontaneous breaking of translation
symmetry:\cite{ReadSachdev90} the dimerized ground state is twofold degenerate
for $n_{c}=2 \bmod 4$ and fourfold degenerate for $n_{c}=(1 \text{\ or\ } 3)
\bmod 4$.

When the $SU(N)$-breaking perturbation $\gamma$ is switched on, the monopoles are
transformed in a similar way as in $d=1$ case for skyrmions: at
$\gamma<0$ monopoles with odd $\widetilde{q}$ are strongly suppressed for $N=3$
and remain unaffected for $N\geq4$. 
Suppression of odd-charged monopoles for $N=3$ and $\gamma<0$ can be understood
by invoking the same type of argument as that we have used in the
one-dimensional case. The monopole charge $\widetilde{q}=\oint j_{\alpha}
dS_{\alpha}$ can be defined as the quantized flux of the ``skyrmion current''
\begin{equation} 
\label{current-CP2} 
j_{\alpha}=\frac{1}{2\pi} \varepsilon_{\alpha\mu\nu}\frac{\partial
  A_{\nu}}{\partial x_{\mu}} = -\frac{i}{2\pi}
  \varepsilon_{\alpha\mu\nu}(\partial_{\mu}\vec{z}^{*}\cdot \partial_{\nu}\vec{z})
\end{equation}
through a closed surface surrounding the monopole.  For antiferromagnetic-type
 configurations favored at $\gamma<0$ one again can write
 $\vec{z}=\frac{1}{\sqrt{2}}(\vec{e}_{1}+\vec{e}_{2})$, where $\vec{e}_{1,2}$
 are two orthonormal vectors, and define the corresponding $O(3)$ unit vector
 field $\vec{\ell}=\vec{e}_{1}\times \vec{e}_{2}$. Then it is easy to obtain
\begin{eqnarray} 
\label{j=2J} 
j_{\alpha}&=&\frac{1}{2\pi}
\varepsilon_{\alpha\mu\nu}(\partial_{\mu}\vec{e}_{2}\cdot
\partial_{\nu}\vec{e}_{1})\nonumber\\
&=&\frac{1}{2\pi}\varepsilon_{\alpha\mu\nu}[\vec{e}_{1}\cdot (\vec{e}_{2}\times
\partial_{\mu}\vec{e}_{2})] [\vec{e}_{2}\cdot (\vec{e}_{1}\times
\partial_{\nu}\vec{e}_{1})]\nonumber\\
&=&\frac{1}{4\pi}\varepsilon_{\alpha\mu\nu}
\vec{\ell}\cdot(\partial_{\mu}\vec{\ell}\times\partial_{\nu}\vec{\ell})
\equiv 2J_{\alpha},
\end{eqnarray}
where $J_{\mu}$ is the corresponding skyrmion current of the $O(3)$
nonlinear sigma model whose flux through a closed surface should be an
integer number. Again, this argument only works for $N=3$. For $N\geq
4$ the $\widetilde{q}=1$ monopole solution\cite{MurthySachdev90}
\begin{equation} 
\label{q1-monop} 
\vec{z}=\vec{U}\cos(\theta/2)e^{i\varphi}+\vec{V}\sin(\theta/2),
\end{equation}
where $\theta$ and $\varphi$ are the angular spherical coordinates in the
(2+1)-dimensional space, and the monopole is assumed to be placed at the origin,
can be easily adjusted to yield $\vec{z}^{2}=0$ identically for $N\geq 4$, and
for $N=3$ the excess action due to the perturbation $\gamma$ of such a monopole
diverges as the spacetime volume (note that this contribution arises due to
deviation of $\vec{z}^{2}$ from $0$ and thus is not destroyed by the vanishing
spin stiffness in the disordered phase as the contribution from the main
$\mathcal{A}_{0}$ part of the action does \cite{MurthySachdev90}).  Even-charged
monopoles can be shown to survive for a finite $\gamma<0$ as exact solutions
(see Appendix \ref{app:A}).  So, we come to the conclusion 
that \emph{at $\gamma<0$ the odd-charged monopoles get confined 
into pairs for $N=3$, but are insensitive to the perturbation for $N\geq
4$}. The consequence for $N=3$ is that the contribution of odd-charged monopoles is
switched off for any finite $\gamma<0$, which effectively amounts to doubling
$n_{c}$ in (\ref{AB-2d}); for the bilinear-biquadratic $S=1$ model (\ref{ham1})
that means that the dimerized state is doubly degenerate at $\gamma<0$ and 
becomes fourfold degenerate only at $\gamma=0$.

On the nematic side ($\gamma>0$) 
the effective theory has been constructed by
Grover and Senthil;\cite{GroverSenthil07} they have shown that the problem can
be mapped to an XY model with a fourfold anisotropy term. The dimerized ground
state is respectively predicted to be fourfold degenerate in that case. One is
thus led to conclude that $\theta=-\pi/2$ for $N=3$ should be the
first order transition line.

\section{Effect of the $SU(N)\mapsto SU(N-1)$ perturbation}
\label{sec:SUN-SUN1}

Consider now a different way to perturb the $SU(N)$ symmetry, namely let us
introduce a finite mass for one of the components of the $\vec{z}$ field,
\begin{equation} 
\label{AD} 
\mathcal{A}\mapsto\mathcal{A}+ \frac{m_{0}^{2}}{2g}\int d^{d+1}x  |z_{N}|^{2},
\end{equation}
which for the $S=1$ model (\ref{ham1}) is equivalent to including the
\emph{easy-axis} single-ion anisotropy term (\ref{hamD}) with
$D=2gm_{0}^{2}$. For cold atoms in optical lattices, such terms appear naturally
in presence of external magnetic field due to the quadratic Zeeman
coupling.\cite{ImambekovLukinDemler03,Zhou+04}
This perturbation breaks the $SU(N)$ symmetry of the model down
to $SU(N-1)$ and produces a $CP^{N-2}$ model with the topological angle $\Theta=\pi$
as the effective theory.  
Actually, the operators
$N_{\pi}^{zz}=\sum_{\vec{n}}\eta_{\vec{n}} t^{\dag}_{\vec{n},N}
t^{\vphantom{\dag}}_{\vec{n},N} $ 
and $N_{0}^{zz}=\sum_{\vec{n}} t^{\dag}_{\vec{n},N}
t^{\vphantom{\dag}}_{\vec{n},N} $
commute with
the Hamiltonian (\ref{ham2}) if  $\cos\theta=0$ and $\sin\theta=0$, respectively.
So, at the ``ferro-$SU(3)$'' point  $\theta=-5\pi/4$ the single-ion
anisotropy $D$ acts simply as an ``external field'' coupling to a
conserved quantity, but at the ``AF-$SU(3)$'' point $\theta=-\pi/2$ the
situation is different.

\begin{figure}[tb]
\includegraphics[width=0.38\textwidth]{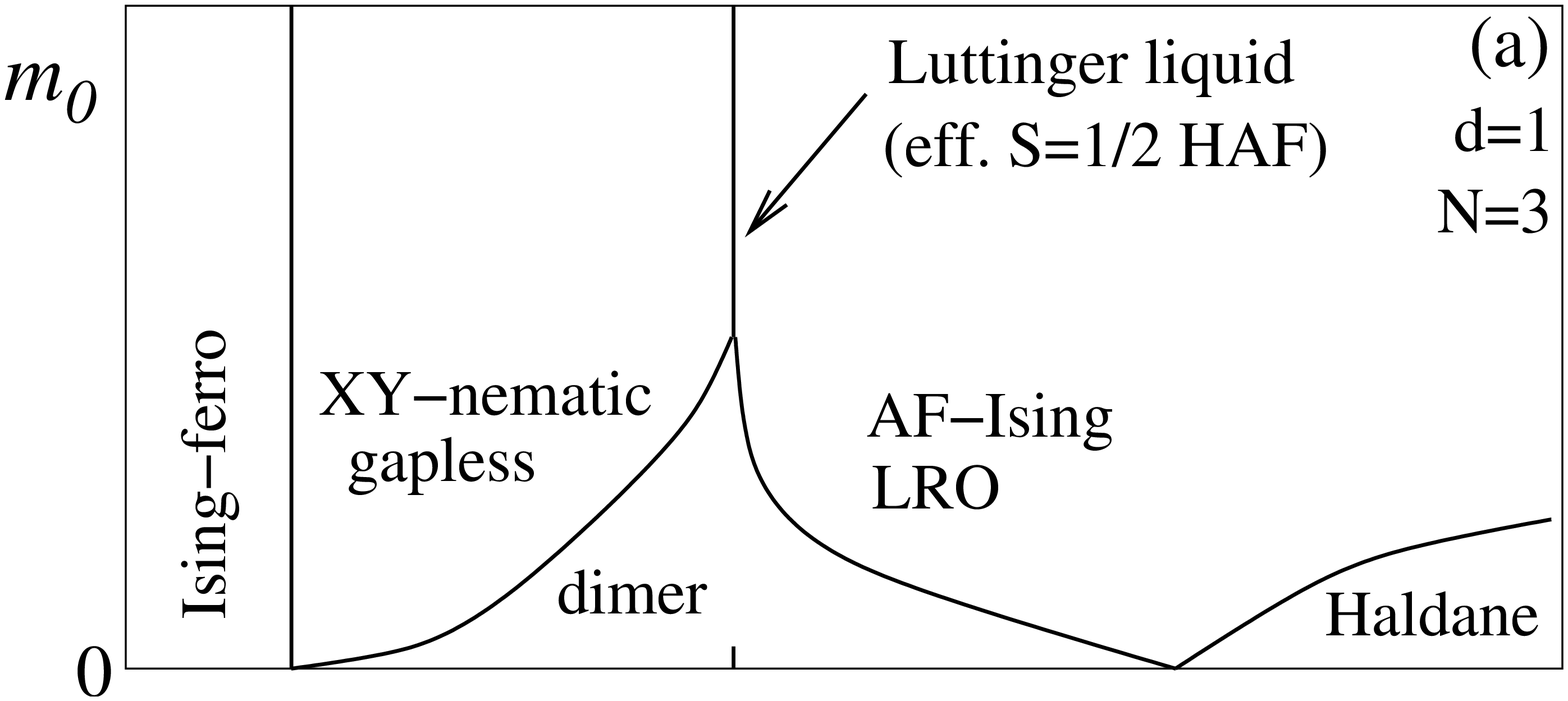}

\includegraphics[width=0.38\textwidth]{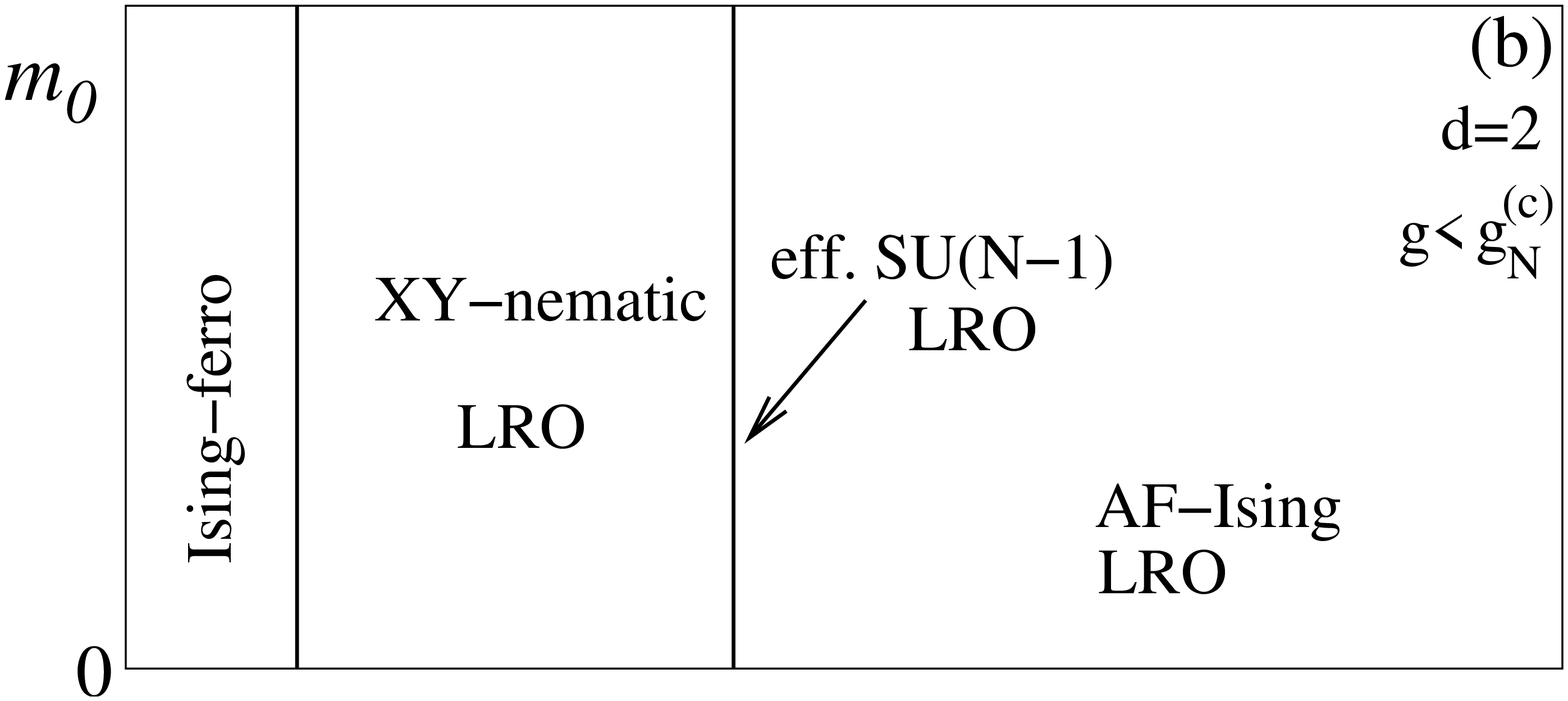}

\includegraphics[width=0.38\textwidth]{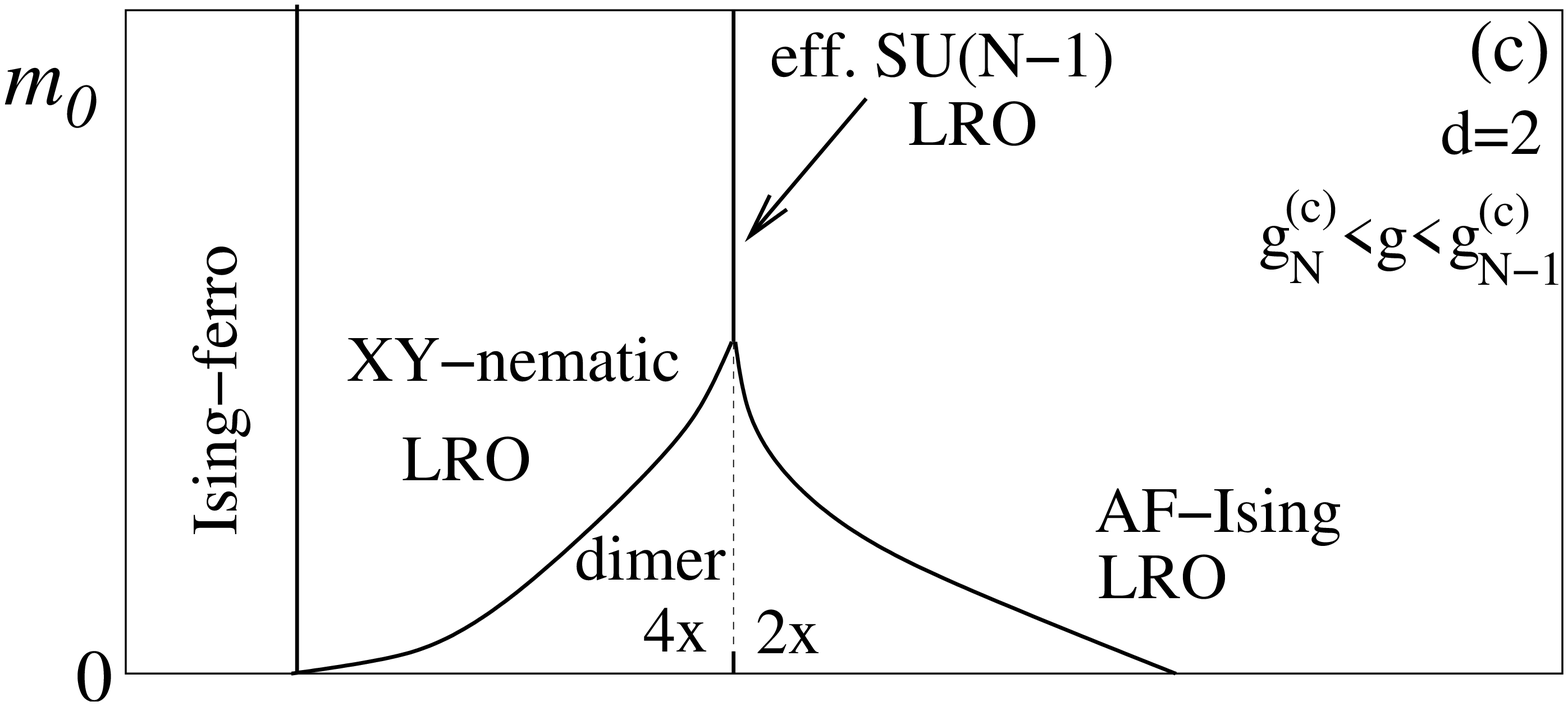}

\includegraphics[width=0.38\textwidth]{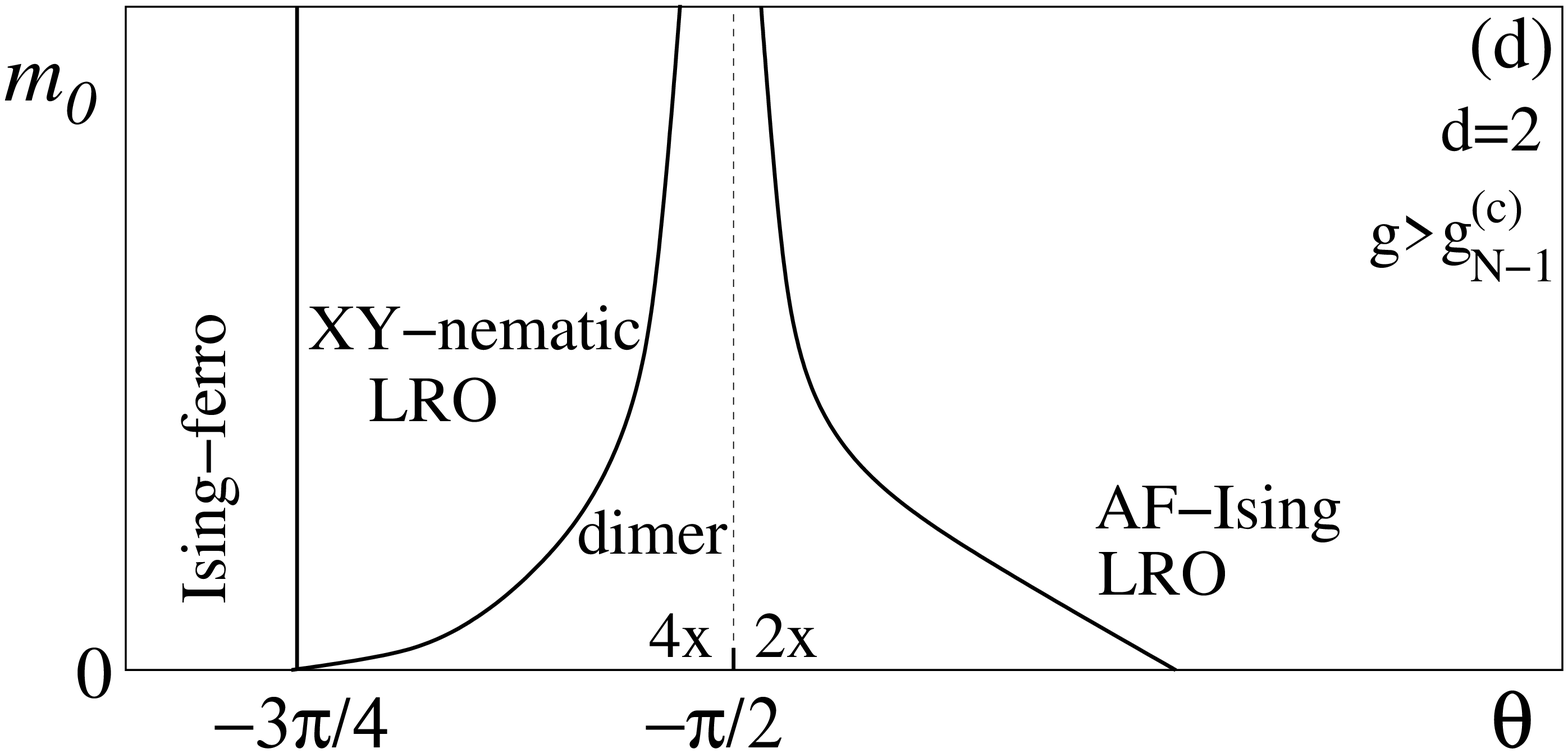}

\caption{\label{fig:DTh-diag}
A sketch of the phase diagram of the model (\ref{ham2}) in the vicinity of the
$SU(N)$-symmetric point $\theta=-\pi/2$, in presence of two symmetry-breaking
perturbations (\ref{Apert}) and (\ref{AD}): (a) the one-dimensional case with
$N=3$; for $N\geq 4$ only the dimerized phase survives around the
$SU(N)$-symmetrical point; (b-d) the
two-dimensional case for different values of the bare coupling
$g=n_{c}^{-1}\sqrt{1+\lambda^{-1}}$; here $g_{N}^{(c)}=\pi^{2}/[\Lambda(N-1)]$;
for $N=3$ only, the degeneracy of the dimerized phase changes from fourfold at
$\theta<-\pi/2$ to
twofold   at $\theta>-\pi/2$.
}
\end{figure}

In one dimension ($d=1$), if the mass $m_{0}$ is large compared to the
gap $\Delta\simeq \Lambda \exp\{-\pi/ g(N-1)\}$, one can integrate out
just the single most massive $N$-th component and obtain in that way a
correspondence between the \emph{bare} coupling constant $g_{N-1}$ of
the effective $CP^{N-2}$ model and the bare coupling constant $g\equiv
g_{N}$ of the original $CP^{N-1}$ model:
\begin{equation} 
\label{geff-N-1} 
g_{N-1}=
\frac{g}{1-\frac{g}{2\pi}\ln\big(1+\frac{\Lambda^{2}}{m_{0}^{2}}\big)}.
\end{equation}
Now, the case $N=3$ is again exceptional because the $CP^{1}$ model
with the topological angle $\Theta=\pi$ in $d=1$ is gapless in an
extended range of coupling.  At infinite coupling $g_{2}=\infty$ the
parity is broken,\cite{Seiberg84} and several
approaches\cite{Affleck91-prl,Azcoiti+07} indicate that there is a
parity-breaking dimerization transition at a strong but finite value
of $g_{2}$, although it seems the answer may depend on the specific
lattice realization.\cite{ShankarRead90} Thus, at least for some range
of $g\equiv g_{3}$ the coupling $g_{2}$ will flow to zero and one
expects a phase transition for $N=3$ on the way from $m_{0}=0$ to
$m_{0}=\infty$. For $N\geq4$ the resulting $CP^{N-2}$ model with
$\Theta=\pi$ remains dimerized, so no phase transition takes place.

In the two-dimensional case, a usual poor-man's RG calculation yields
the effective coupling $\widetilde{g}$ as a function of the anisotropy $m_{0}^{2}$: 
\begin{equation}
\label{geff-N-2}
\widetilde{g}=
\frac{g}{1-\frac{g}{2\pi^{2}}
\big[(N-1)\Lambda -m_{0}\arctan\frac{\Lambda}{m_{0}}\big]}.
\end{equation}
Depending on the value of the bare coupling
$g=n_{c}^{-1}\sqrt{1+1/\lambda}$, 
there are three
possible scenarios: (a) if $g<g_{N}^{(c)}=\pi^{2}/[\Lambda(N-1)]$,
then the system has long-range nematic or AF order all the way from $m_{0}=0$ to
$m_{0}=\infty$; (b) if $g_{N}^{(c)}< g < g_{N-1}^{(c)}$,  then the
system is disordered (and dimerized) at $m_{0}=0$, but with increasing
$m_{0}$ there is an ordering transition at $m_{0}\simeq
(2\pi/g_{0})(1-g/g_{N}^{(c)})$;
finally, if $g>g_{N-1}^{(c)}$, the system stays dimerized at all
values of $m_{0}$. 
The combined effect of the $SU(N)\mapsto SU(N-1)$-breaking
perturbation (\ref{AD}) and the $SU(N)\mapsto O(N)$ one (\ref{Apert})
is also transparent: taken together, those terms lower the symmetry to
$O(N-1)$, and for $N\geq 4$ the corresponding behavior as a function
of $\gamma$ at finite
$m_{0}$ can be inferred
from the behavior of the model with $N\to N-1$.
In one dimension, for $N=3$ and at large $m_{0}$,  $\gamma>0$
favors a phase with dominant power-law $XY$-type nematic correlations
(the XY2 phase in the classification of Schulz \cite{Schulz86}),
while $\gamma<0$ favors the Ising-type long-range antiferromagnetic order. The
transition from the XY-nematic to the dimerized phase is of the
Berezinskii-Kosterlitz-Thouless type, and the transition from the
AF-Ising to the dimerized phase belongs to the Ising universality class.
The corresponding phase diagrams are sketched in Fig.\
\ref{fig:DTh-diag}.

For the spin-$1$ model (\ref{ham1}) that corresponds to $N=3$, it is instructive
to construct the effective Hamiltonian in the limit of strong
single-ion anisotropy $D\gg 1$.  Indeed, in that limit the Hilbert
space at each site $\vec{n}$ is effectively reduced to the two spin-$1$ states
$|+\rangle$, $|-\rangle$, which can be identified with
$|\!\uparrow\rangle$ and $|\!\downarrow\rangle$ states of 
pseudospin-$\frac{1}{2}$.   In the
second order of perturbation theory in $1/D$, the effective 
Hamiltonian  is given by the XXZ model in terms of pseudospin-$\frac{1}{2}$ 
operators $\vec{\tau}_{\vec{n}}$:
\begin{eqnarray} 
\label{Heff-D} 
\mathcal{H}_{\rm eff}&=&\sum_{\vec{n}} \Big\{ \widetilde{h}_{\vec{n},\vec{n}+\vec{x}}
+\lambda\widetilde{h}_{\vec{n},\vec{n}+\vec{y}} \Big\},\nonumber\\
\widetilde{h}_{\vec{n},\vec{n'}}&=&-\widetilde{J}_{xy}(\tau_{\vec{n}}^{x}\tau_{\vec{n}'}^{x} + 
\tau_{\vec{n}}^{y}\tau_{\vec{n}'}^{y}) 
+ \widetilde{J}_{z}\tau_{\vec{n}}^{z}\tau_{\vec{n}'}^{z},\\
\widetilde{J}_{xy}&\simeq& -2\sin\theta +\frac{(\cos\theta-\sin\theta\sqrt{2})^{2}}{2D},\nonumber\\
\widetilde{J}_{z}&=&\widetilde{J}_{xy}+4\cos\theta.\nonumber
\end{eqnarray}
One can see that for $\theta=-\pi/2$ the effective Hamiltonian is
$SU(2)$-symmetric, in agreement with the continuum field
description. Deviations from $\theta=-\pi/2$ break this $SU(2)$ symmetry,
favoring AF or nematic order.

\section{Summary}
\label{sec:summary}

We have studied the consequences of explicit symmetry breaking in the
model of low-dimensional $SU(N)$ antiferromagnet on a bipartite
lattice, motivated by the
physics of cold spinor bosonic atoms in optical lattices. Two possible
routes have been considered: lowering the $SU(N)$ symmetry down to
$O(N)$ and to $SU(N-1)$. Physically, in cold atom systems those perturbations naturally
arise due to the presence of the external magnetic field which
controls the detuning from the Feshbach resonance and simultaneously
causes the quadratic Zeeman effect. Both ways of the symmetry breaking
result in rich sequences of transitions between dimerized,
antiferromagnetic, and spin-nematic phases. The qualitative form of
the phase diagram depending on the model parameters is established.
It is shown that the physically interesting case $N=3$ is special:
perturbing the $SU(3)$ symmetry leads to nontrivial changes in the
Berry phases, which are reflected in the degeneracy of
the dimerized phase.

\section*{Acknowledgments}

I would like to thank Boris Ivanov, Ian McCulloch, Subir
Sachdev and Kun Yang for stimulating discussions.  The present work
has been
supported by the Heisenberg Program Grant No.\ KO~2335/1-2 from
Deutsche Forschungsgemeinschaft.

\appendix
\section{Even-charged skyrmions and monopoles in the perturbed $CP^{2}$ model}
\label{app:A}

Consider a general $q=2$ skyrmion solution of the unperturbed  ($\gamma=0$)
(1+1)-dimensional $CP^{2}$ model, which has the form
\begin{equation} 
\label{q2-gen} 
z_{\alpha}=f_{\alpha}/|\vec{f}|, \quad
f_{\alpha}=a_{\alpha}Z^{2}+b_{\alpha}Z+c_{\alpha}.
\end{equation}
Now we would like to adjust the parameters of
the above solution to satisfy $\vec{z}^{2}=0$, making it suitable for $\gamma<0$. Denoting the real and
imaginary parts of the three-component complex vectors $\vec{a}$, $\vec{b}$, and $\vec{c}$ as
$\vec{a}_{1}$, $\vec{a}_{2}$ etc., we see that $(\vec{a}_{1}, \vec{a}_{2})$ 
 must be  (up to a scale factor) a pair of mutually orthogonal unit
vectors, and the same is true for  $(\vec{c}_{1}, \vec{c}_{2})$. We choose the
coordinate system so that
$\vec{c}_{1}\parallel \widehat{\vec{x}}$ and $\vec{c}_{2}\parallel
\widehat{\vec{y}}$, and set $b_{z}=2R>0$  to fix the overall phase and norm. The
following solution does the job:
\begin{eqnarray} 
\label{q2-af} 
&&\vec{b}=\{ \zeta^{*},\zeta,2R\},\quad \vec{c}=R^{2}\{1,i,0\},\quad 
\vec{a}=\vec{a}_{1}+i\vec{a}_{2},\nonumber\\
&& \vec{a}_{1}=\big\{ 
-1-\lambda^{2}\cos 2\chi , \lambda^{2}\sin 2\chi, 2\lambda\sin\chi
\big\},\nonumber\\
&& \vec{a}_{2}=\big\{ 
-\lambda^{2}\sin 2\chi, 
1-\lambda^{2}\cos 2\chi,
-2\lambda\cos\chi
\big\},
\end{eqnarray}
where $\zeta$ is an arbitrary complex number and $\lambda$, $\chi$ are
real. It is easy to convince oneself that this solution is nothing but
the disguised Belavin-Polyakov skyrmion\cite{BelavinPolyakov75} of the $O(3)$
nonlinear sigma model with the topological charge $Q=1$. The
correspondence between the $CP^{2}$ field $\vec{z}$ and the
sigma-model unit vector $\vec{\ell}$ is given by
$\vec{\ell}=-i(\vec{z}^{*}\times \vec{z})$, and the $O(3)$ topological
charge is determined by (\ref{q=2Q}).  One can easily see that the
simplest Belavin-Polyakov solution
$
(\ell_{1}+i\ell_{2})/(1-\ell_{3})=Z/R
$
translates into
\[
\vec{z}=\frac{1}{\sqrt{2}(|Z|^{2}+R^{2})}\big\{ Z^{2}+R^{2}, i(R^{2}-Z^{2}), 2iRZ\big\}
\]
which after a rotation $Z\mapsto Ze^{-i\pi/2}$ becomes a special case of
(\ref{q2-af}) with $\lambda=0$, $\chi=\pi/2$, and $\zeta=0$. This solution
describes a $q=2$ skyrmion whose six constituents (``zindons'') sit at $Z=\pm
R$, $Z=\pm iR$, $Z=0$ and $Z=\infty$.

In a similar way, one can show that in (2+1) dimensions a monopole of the
$CP^{2}$ model with the even integer charge $\widetilde{q}=2m$, defined as a
solution to the equation \cite{Polyakov-book}
\begin{equation} 
\label{eq-monop-CP2}
\varepsilon_{\alpha\mu\nu} \frac{\partial A_{\mu}}{\partial x_{\nu}}
=\frac{\widetilde{q} x_{\mu}}{2r^{3}},
\end{equation}
where $r^{2}=\sum_{\mu} x_{\mu}^{2}$ and the monopole is assumed to be at the
origin, corresponds exactly to the hedgehog solution of the $O(3)$ model with a
charge $\widetilde{Q}=m$. Indeed, it is straightforward to check that the
solution of the form
\begin{equation} 
\label{monop-CP2} 
\vec{z}=\frac{1}{\sqrt{2}}\left\{
\begin{array}{c}
\cos\theta\cos(m\varphi)-i\sin(m\varphi)\\  
\cos\theta\sin(m\varphi)+i\cos(m\varphi)\\
 -\sin\theta
\end{array}
\right\},
\end{equation}
where $\theta$ and $\varphi$ are the spherical angular coordinates in
the (2+1)-dimensional space, satisfies (\ref{eq-monop-CP2}) with $\widetilde{q}=2m$,
satisfies $\vec{z}^{2}=0$,
and its corresponding  $O(3)$ unit vector field $\vec{\ell}=-i(\vec{z}^{*}\times
\vec{z})$ describes a $\widetilde{Q}=m$ hedgehog:
\begin{equation} 
\label{monop-O3} 
\vec{\ell}=\big\{ 
\sin\theta\cos\varphi, \sin\theta\sin\varphi,\cos\theta
\big\}. 
\end{equation}
This confirms that the even-charged monopoles (\ref{monop-CP2}) remain exact
solutions even in the perturbed case (but only for  $\gamma<0$). Odd-charged
monopoles are suppressed as explained in Sect.\ \ref{subsec:Berry}.



\begin{thebibliography}{10}
\bibitem{comment1} The effective theory of Ref.\ \protect\onlinecite{IvanovKolezhuk03}
  has to be augmented with the Berry phase contributions coming from
  disclinations of the director field, yielding the picture similar to that
  presented in Ref.\ \protect\onlinecite{GroverSenthil07} (see Sect.\ \protect\ref{subsec:Berry}).
\end{thebibliography}
\end{document}